# A 12-mode Universal Photonic Processor for Quantum Information Processing

C. Taballione[1], R. van der Meer[2], H.J. Snijders[1], P. Hooijschuur[2], J.P. Epping[1], M. de Goede[1], B. Kassenberg[1], P. Venderbosch[1], C. Toebes[2], H. van den Vlekkert[1], P.W.H. Pinkse[2], and J.J. Renema[1,2]

[1] *QuiX BV, Hengelosestraat 500, 7521 AN Enschede, The Netherlands*
[2] *MESA+ Institute for Nanotechnology, University of Twente, PO Box 217, 7500 AN Enschede, The Netherlands*

**Abstract:** Photonic processors are pivotal for both quantum and classical information processing tasks using light. In particular, linear optical quantum information processing requires both large-scale and low-loss programmable photonic processors. In this paper, we report the demonstration of the largest universal quantum photonic processor to date: a low-loss, 12-mode fully tunable linear interferometer with all-to-all coupling based on stoichiometric silicon nitride waveguides.

## 1. Introduction

Photonic processors, also called Universal Multiport Interferometers (UMI) or Photonic FPGAs, have attracted increasing attention in the past years for their many fields of applications such as quantum information processing based on linear optics [1-13], quantum repeater networks [14-17], (quantum) machine learning [18-20] and radio-frequency signal processing [21, 22]. A photonic processor is a tunable multimode interferometer which can achieve arbitrary optical transformations. Various realizations have been proposed in literature, where photonic processors have been arranged in many different topologies: triangular [23, 1], rhomboidal [4], fan-like [19], square [24], hexagonal [22] and quadratic [21].

Linear optics quantum information processing holds great promise for solving particular problems with exponentially greater computational power than classical computers. A large collection of proposed applications can be found in the literature [25-27]. The recent a demonstration of a quantum advantage in a static optical system [28] shows the urgent need for programmable photonic processors.

The fundamental process of linear optics quantum information processing is quantum interference. To exploit it, a setup is needed consisting of photon sources, a photonic processor and single-photon detectors. The photons are used as information carriers and the photonic processor, formed by linear optical elements, will process quantum information by letting the photons interfere in a controlled manner. By looking at the configurations of the output samples of the photons at the detectors, the result of the photonic computation can be read out.

For photonic quantum information processing, the requirements on a photonic processor are fourfold. First, it must be large-scale as this increases the complexity of the problems that can be solved. Second, it must be universal (i.e., fully-reconfigurable and with all-to-all connectivity) since this enables the implementation of arbitrary transformations mapping onto various problems. Third, it must be low-loss as otherwise the information carried by single photons is lost. Finally, as stated above, a photonic processor needs to preserve quantum interference.

In this paper, we describe a 12-mode[i] low-loss (end-to-end 5 dB) reconfigurable photonic processor based on stoichiometric silicon nitride waveguides, which is the largest universal photonic processor to date. We report the results of the classical and quantum characterization showing that full reconfigurability, low loss, and high-fidelity operations are achieved.

The paper is structured as follows: in section 2 the components of the 12-mode photonic processor are described; in section 3 we report the experimental results of the classical and quantum characterization of the processor; in section 4 prospects for the technology are discussed; in section 5 we derive the conclusions.

### 2. Photonic processor

In this section, we describe the main components of our 12-mode photonic processor (Fig. 1). It consists of three parts: an integrated silicon nitride photonic chip, peripheral equipment, and the software to control its functionality.

### Photonic chip

The heart of our photonic processor is a reconfigurable photonic integrated circuit based on stoichiometric silicon nitride ($Si_3N_4$) waveguides with the TripleX technology. Thanks to the chosen material platform, we achieve propagation losses as low as 0.1 dB/cm with a minimum bending radius of 100 μm. The waveguide cross-section used in the photonic processor is an asymmetric double-stripe (ADS) [29] as shown in Fig. 1a. The waveguides are designed for single-mode operation at a wavelength of 1550 nm. ADS waveguides enable low-loss coupling to standard telecom fibers using spot-size

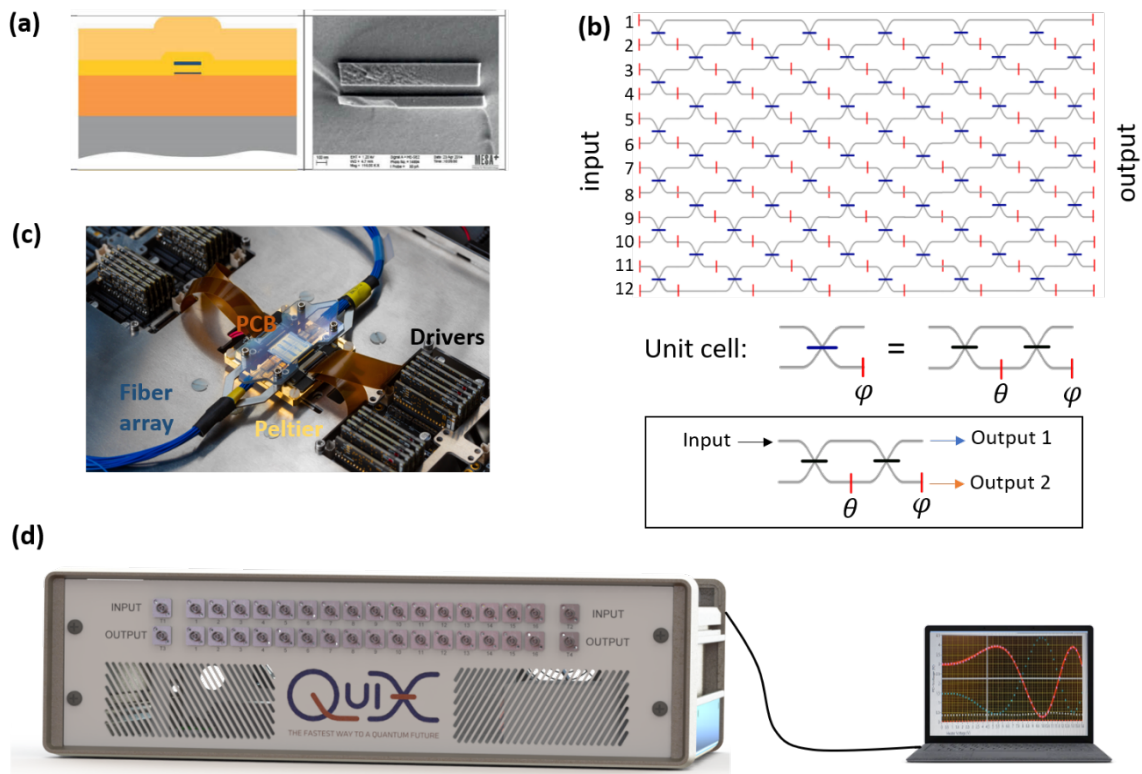

Figure 1 Overview of the QuiX photonic processor. (a) Schematic and SEM picture of the asymmetric double-stripe cross-section used for the waveguides in this paper. (b) Functional design of the 12-modes photonic processor. The blue line represents a TBS that is implemented as a MZI with two 50:50 directional couplers (black lines) and a thermo-optic phase shifter in red. When calibrating the unit cell, light is injected, for example, in the top input while both outputs are monitored. (c) Picture of the photonic assembly of the 12-modes processor as mounted inside the control box. (d) Schematic for the QuiX control system, i.e., the control box is remotely controlled via a software interface in Python.

converters, where the upper silicon nitride stripe is removed by adiabatic tapering [29].

The reconfigurability of the photonic processor is achieved by exploiting the thermo-optic via resistive heating of 1 mm-long platinum phase shifters. Using these thermo-optic phase shifters, a $\pi$ phase shift is achieved at $V_\pi \cong 10$ V, corresponding to an electrical power of ~385 mW per element.

The functional design of our processor is presented in Fig. 1b. An optical unit cell, composed of a tunable beam splitter (blue line) and a phase shifter (red line) on the bottom output mode, is repeated 66 times over a square topology of 12 input/output modes and circuit depth of 12. Twenty-four additional phase shifters are distributed across the inputs and outputs for sub-wavelength delay compensation and external phase tuning. In total, the processor contains 156 phase shifters (of which 24 are redundant and not connected). The tunable beam splitters are implemented by Mach-Zehnder interferometers (MZI) consisting of two 50:50 directional couplers (black lines in Fig. 1b) and an internal phase shifter, θ, followed by an external phase shifter, ϕ, at the bottom output mode. Each unit cell represents a node of the large-scale interferometer where light can interfere.

**Peripherals**

The photonic processor presented above is embedded in a control box that includes electronic and temperature control modules (Fig. 1c and 1d). The reconfigurable photonic integrated circuit is optically packaged with polarization-maintaining (PM) fiber arrays for the in- and out-coupling of light. For ease of access the input and output fibers are fixed to the front panel of the control box via PM mating sleeves. We measure an average loss for the 24 PM mating sleeves of about 0.18 dB/connector.

A printed circuit board (PCB) was fabricated and wire-bonded to the photonic processor. A total of 132 voltage drivers are connected to the PCB enabling the independent tuning of each thermo-optic phase shifter, achieved via serial communication with a standard pc.

Temperature control and stability of the processor is achieved by active cooling. A thermo-electric Peltier element is placed beneath the sub-mount of the packaged processor, i.e., a metallic mount holding the processor, the fiber arrays and the PCB. The Peltier element favors the heat transfer in the vertical direction, from the on-chip phase shifters to the heat-sink. To further increase the heat capacity of the system, water cooling can also be installed.

**Software – Control system**

The optical transmission through our photonic processor is described by a matrix, *S*, relating the output electric fields to the input ones $E_{out} = S E_{in}$. The matrix *S* is given by the product of two-mode matrices $S_{m,n}$ of each unit cell between mode *m* and *n*. Considering a unit cell as in Fig. 1b with pairs of ideal and symmetric 50:50 directional couplers (k = 0.5), we find that

$$S_{unit\_cell}(\theta, \phi) = \frac{1}{2}\begin{pmatrix} 1 - e^{-i\theta} & -i(e^{-i\theta} + 1) \\ -i(e^{-i\theta} + 1) \cdot e^{-i\phi} & -(1 - e^{-i\theta}) \cdot e^{-i\phi} \end{pmatrix}$$

$$S_{m,n}(\theta, \phi) = e^{-i(\theta - \pi)/2} \cdot$$

$$\cdot \begin{pmatrix} 1 & 0 & \cdots & \cdots & \cdots & 0 \\ 0 & \ddots & & & & \vdots \\ \vdots & & \sin\frac{\theta}{2} & -\cos\frac{\theta}{2} & & \vdots \\ \vdots & & -\cos\frac{\theta}{2} \cdot e^{-i\phi} & -\sin\frac{\theta}{2} \cdot e^{-i\phi} & & \vdots \\ \vdots & & & & \ddots & 0 \\ 0 & \cdots & \cdots & \cdots & 0 & 1 \end{pmatrix}$$

where θ is the internal phase of the MZI and ϕ is the external phase. By varying θ and ϕ over $2\pi$ it is possible to perform any transformation in the special unitary group, SU(2). It is important to note that the action of the processor is always described by the same classical scattering matrix S independently of the nature of the input state, i.e., either classical electric field amplitudes or photon-number states. Therefore, to know the transmission matrix, it is sufficient to characterize the processor classically. In the case of quantum input light, the formalism becomes $\hat{a}_{out} = S\hat{a}_{in}$ where now the scattering matrix relates the ladder operators instead of classical electric field amplitudes [30].

The combination of quadratically many $S_{m,n}$ allows the implementation of any complex-valued unitary transformations U. We can decompose any arbitrary unitary transformation U into sets of (θ, ϕ)$_{m,n}$ belonging to specific unit cells between mode m and n of our processor [23, 24]. Assigning these (θ, ϕ)$_{m,n}$ to the corresponding scattering matrix $S_{m,n}$ and multiplying them in the order of light propagation through the processor, the exact optical response corresponding to U will be reproduced.

## 3. System performance

In this section, we report the classical and quantum characterization of the 12-mode photonic processor. The experimental setup is depicted in Fig. 2.

For classical characterization of the processor we use a CW diode laser at 1550 nm (2 mW output power– LP1550 PAD). A PM fiber switch can be used to facilitate the procedure of characterization switching the input light across all the 12 inputs. For intensity measurements we use a set of InGaAs photodiodes each mounted on a FC/PC bulkhead (Thorlabs FGA01FC) (Fig. 2a). The output signal of the QuiX hardware, impinging on the PD array, is acquired and read out via an NI BNC-2090A and USB-6211 card.

For quantum characterization a 2 mm ppKTP crystal emitting collinear frequency-degenerate cross-polarized single-photon pairs is pumped with a Ti:Sapphire (Tsunami, Spectra Physics) mode-locked laser with a center wavelength of 775 nm. The photons are measured on superconducting nanowire single-photon detectors (Photon Spot) (Fig. 2b). Coupling between the light source and the QuiX hardware is done via polarization maintaining fibers.

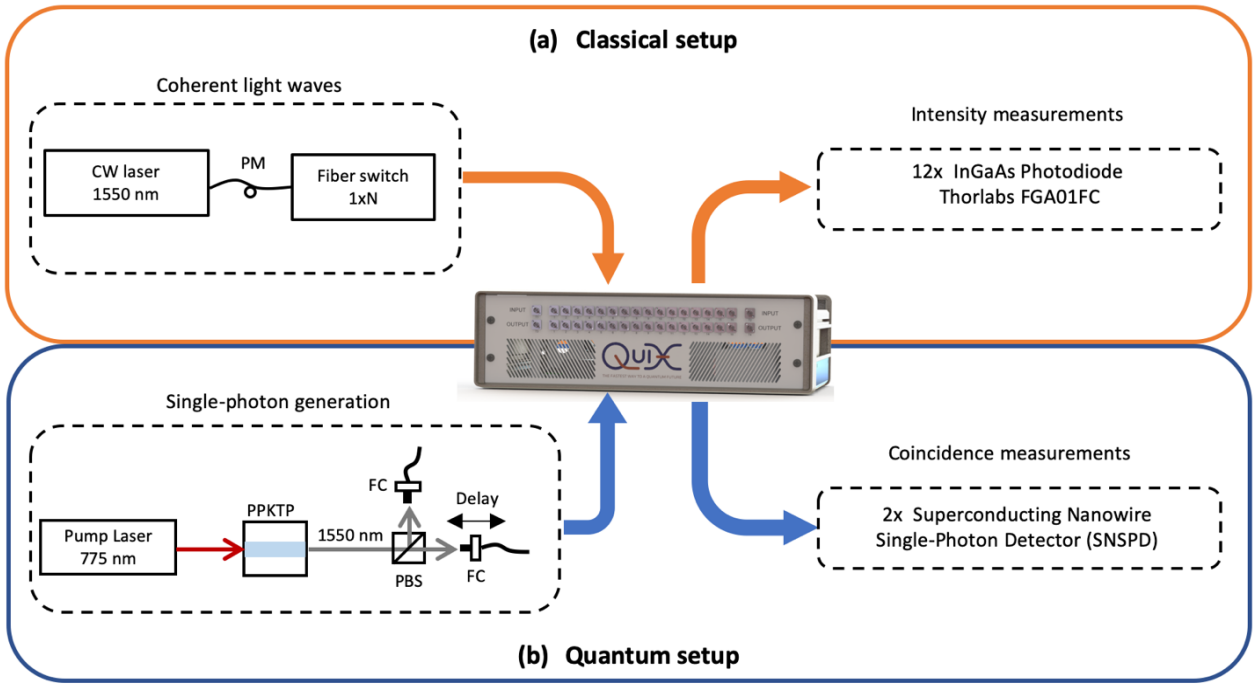

Figure 2 Experimental setup for (a) classical and (b) quantum characterization. (a) CW light is generated by a diode laser at 1550 nm and injected into each input channel of the QuiX system via a PM 1x8 fiber switch. After propagating through the chip, light is detected by an array of 12 photodiodes, one per output channel. (b) A laser at 775 nm pumps a nonlinear PPKTP crystal, generating infrared collinear cross-polarized pair of photons. After polarization separation via a polarizing beam splitter (PBS), each single-photon impinges on a fiber-coupled collector (FC) and is injected into the chip. One of the collectors moves on a translational delay stage to temporally overlap the generated photons. The single photons at the output are detected with two superconducting nanowire single-photon detector (SNSPD) and coincidence measurements are performed.

**Classical response**

First, we report the classical characterization of the processor. This comprises the calibration of all the tunable elements and the total transmission of the processor, as shown in Fig. 3(a, b, c). Specific measurement protocols are used to characterize the tunable beam splitters (TBS) and phase shifters (PS).

In Fig. 3a we show the calibration of one of the on-chip heaters as, for example, the one belonging to a TBS. Both outputs of the TBS are monitored while only the internal phase $\theta$ is varied (inset Fig.1b). This is done by applying a varying voltage, V, to the thermo-optic phase shifter. The same procedure is adopted for characterizing the PSs. By following a specific order, we characterize the entire processor and find that all the 132 thermo-optic phase shifters are tunable over more than $2\pi$ phase range with high extinction ratio. In this way, we achieve full control over the processor.

In Fig. 3b, we report the insertion loss matrix of the processor. The insertion loss matrix includes both the fiber-to-chip-to-fiber coupling losses and the on-chip propagation loss. It excludes thus the aforementioned connector loss of 0.18 dB/connector. The reader can clearly see that there is one input channel that shows higher losses than the others. This is confirmed by inspection of one of the fibers, which turned out to be damaged. The optimal transmission for this input channel can be easily retrieved by exchanging the damaged fiber. On average, the processor shows an insertion loss of ~5 dB where ~0.8 dB comes from propagation loss and the remaining ~4.2 dB are coupling losses.

Finally, to confirm the universality and control of the processor, we perform a large variety of 12-dimensional unitary transformations as summarized in Fig. 3(c, d, e). For each target transformation $U_i$, we measure the corresponding experimental output intensity distribution $|U_{exp}|^2$. We compare the experimental results with the target intensity output distribution $|U_i|^2$ via the fidelity $\mathcal{F}_i = \frac{1}{D} Tr(|U_i^\dagger| \cdot |U_{exp}|)$, where $D$ is the dimension of the unitary transformation and our processor, i.e., $D = 12$.

We perform unitary transformations spanning various applications such as permutation (P) and Haar-random matrices, high-dimensional Pauli-X gates (X) and optical switching matrices (S). Furthermore, to ultimately illustrate our full control over our processor we implement the letters Q and X of our company name QuiX.

In Fig. 3c, we report, as an example, the target (theory) $P_{th}$, $X_{th}$ and $S_{th}$ matrix (top row- from left to right) and their corresponding experimental implementations (bottom row). Fig. 3d shows the distribution of fidelities for 1000 Haar random unitaries. We note that an increase in the complexity of the optical transformation is associated with a decrease in fidelity, as it is natural to expect. Since by definition Haar matrices cover the space of unitary transformations in a uniform way, we expect the fidelity for this set to be representative for an arbitrary transformation. Finally, Fig. 3e reports the measurements of the first and last letter of the company name QuiX. The results are summarized in Table 1.

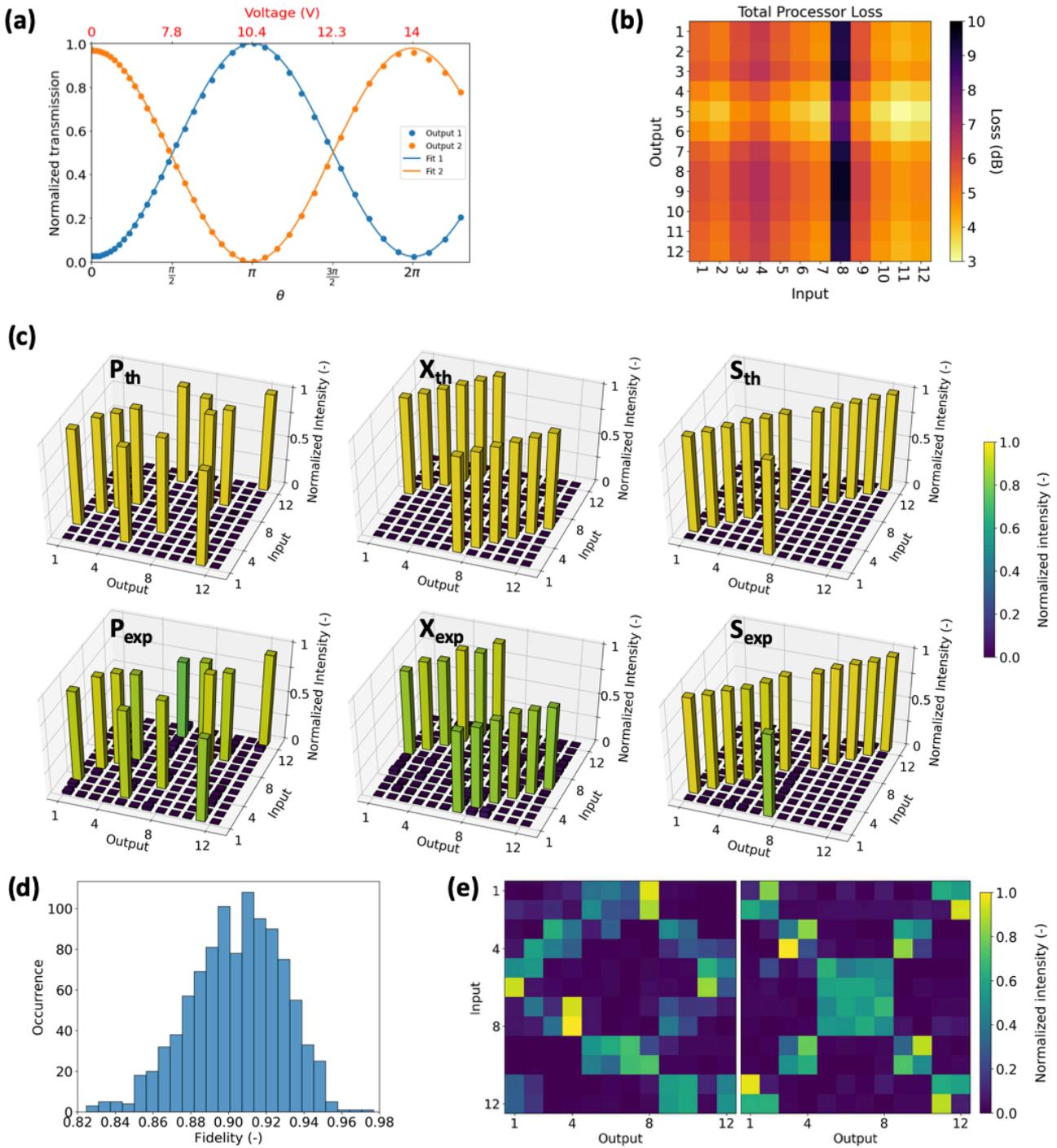

Figure 3 Results summary of the classical characterization. (a) Calibration of heater 136 (TBS). The normalized transmission at the two outputs is plotted and fitted versus the internal phase θ and the applied voltage. A $V_\pi \cong 10.4\ V$ is measured. (b) Setup transmission for each input-output combination. (c) Theory vs experimental realization of one of the random permutation P, $X^6$ and one switching matrix S (between mode 1 and 7) with fidelity, respectively, of 0.955, 0.940 and 0.984. (d) Fidelity distribution of 1000 Haar random matrices. (e) Measured Q and X unitary transformation of fidelity 0.922 and 0.930, respectively. Intensity is normalized to its maximum for each measurement.

| | # | $\mathcal{F}_{ave} \pm \sigma$ |
|---|---|---|
| Random Perm. | 12 | $0.930 \pm 0.013$ |
| Pauli-$X^{n=0,...,12}$ | 12 | $0.945 \pm 0.007$ |
| Switching | 12 | $0.985 \pm 0.006$ |
| Haar Random | 1000 | $0.904 \pm 0.024$ |
| Logo | 2 | $0.926 \pm 0.004$ |

Table 1 Summary of measured fidelities.

## Quantum response

After having demonstrated full control of the processor via the classical response characterization, we attach the QuiX hardware to the quantum light source as shown in Fig. 2b. Quantum interference experiments were performed across the whole photonic processor evaluating to what extent the processor is able to preserve the indistinguishability of the input single photons. This measurement tests all sources of which-way information, such as path-dependent dispersion.

The single-photon source is characterized separately by running a Hong-Ou-Mandel (HOM) [31] interference experiment over a tunable fiber-splitter that gives a visibility of 0.94.

By choosing inputs pairwise, we run Hong-Ou-Mandel (HOM) interference experiments on every single TBS on the processor (Fig. 4). The average visibility of the on-chip HOM interference dips is calculate as $vis_{ave} = \sum_n \frac{\left(1-\frac{cc_{ind}}{cc_{dist}}\right)_n}{n_{tot}}$ where $n_{tot} = 66$, where

$cc_{ind/dist}$ is the coincidence count rate for indistinguishable/distinguishable photons (as indicated in Fig. 4). We obtain an average visibility of $vis_{ave} = 0.923$. We observe from Fig. 4c that the distribution of the HOM visibility is rather random across the UMI confirming the absence of any systematic error in the processor. Some TBSs, e.g. # 136, 140 and 145, present a low visibility of the quantum interfere: this is due to an imperfectly optimized 50:50 splitting ratio setting of these TBSs, which reduces the HOM visibility.

Comparing the reference and the measured average on-chip visibility we can conclude that the processor does not affect the spectral-temporal indistinguishability of the signal and idler photons of the photon pair as coming out from the source. The on-chip visibility is ultimately limited by the source itself.

## 4. Discussion

Finally, we discuss the future prospects of our technology. With this 12 x 12 processor, we have not exhausted the capabilities of the SiN platform; we anticipate producing larger processors with higher fidelity and lower optical loss in the future. We discuss these issues in turn.

The fidelity of the unitary transformations can be improved by correcting and compensating for crosstalk, upgrading both the hardware and the software. Examples can be found in the literature [32-34]. Furthermore, the compensation of non-ideal extinction ratio of the TBS will also improve the fidelity. This can be obtained by redundancy [35, 36].

The insertion loss of the system can be further reduced by optimal waveguide engineering, to enable even a greater scalability of our technology. By optimal waveguide tapering the coupling losses can be reduced down to ~1 dB [29]. With these modifications, the system will become practical as a photonic processor for quantum interference experiments in the regime where a quantum advantage exists [28]. The integration of single-photon sources, by exploiting the third-order nonlinearity of silicon nitride, and detectors will help further in reducing the coupling losses [37-39].

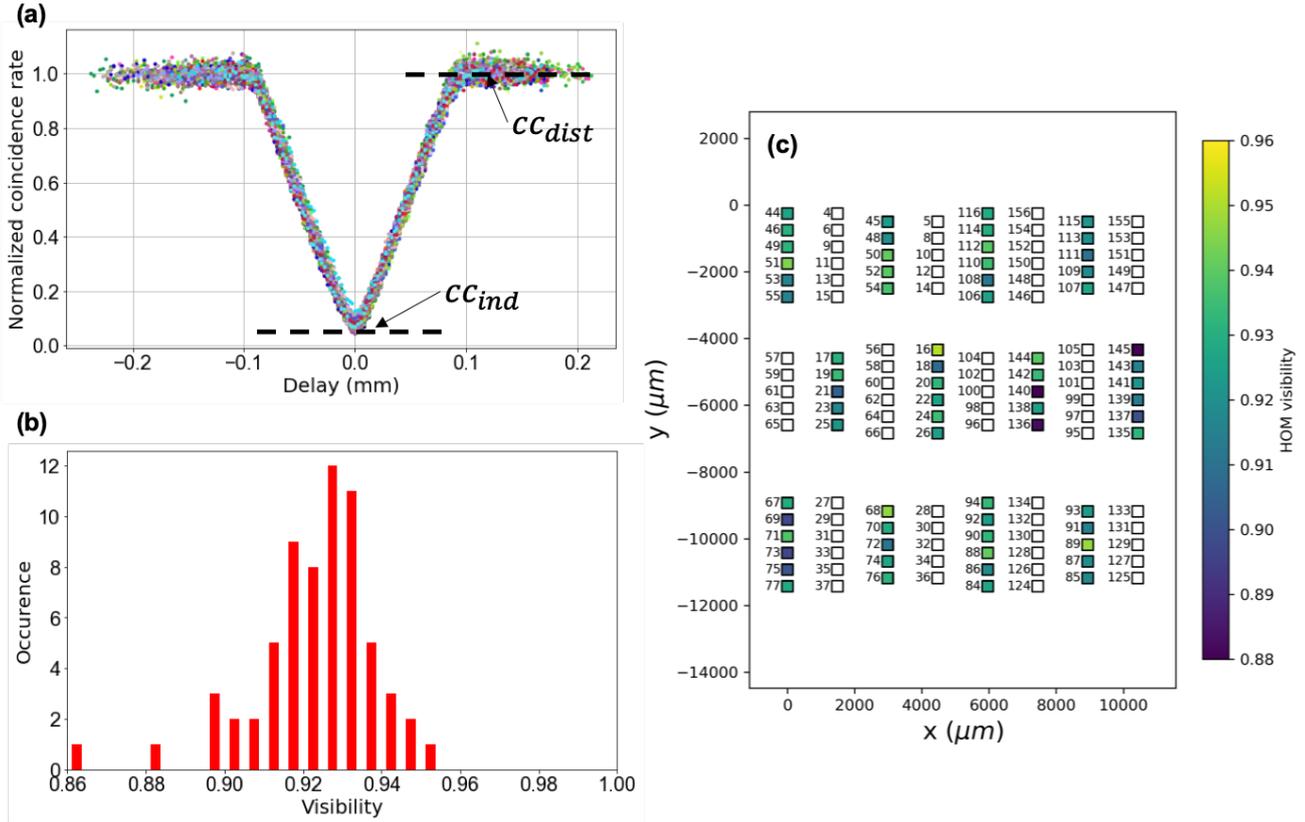

Figure 4 Result summary for the quantum characterization. (a) All the 66 HOM dips are reported on top of each other. The coincidence count rate is normalized to the average of coincidence counts outside the dip, i.e., for distinguishable single photons. (b) Distribution of the visibility of the 66 HOM dips showing an average of 0.923. The two outliers belong to TBS #136 and #140, as can be seen in (c), where the distribution of the visibilities is reported across the heater layout.

Valid alternatives to thermo-optic tuning are available such as the implementation of liquid crystals [40, 41], phase-changing materials [42] and stress-optic tuning [43, 44] in order to reduce the power consumption. The latter additionally has the advantage to operate at cryogenic temperatures [45]. Operating the chip at cryogenic temperatures would permit direct integration with both solid-state single-photon sources [46] and superconducting single-photon detectors [38, 39].

To reduce the foot print of the unit cell, alternative approaches can be undertaken, e.g., by dual-drive directional couplers [47], dual-drive MZI [5] and MEMS [48].

## 5. Conclusions

In this paper we have reported a 12-mode fully reconfigurable universal photonic processor based on silicon nitride waveguides. The processor is embedded into a control system that enables remote access to its optical functionality and reconfigurability. The system operates at 1550 nm with insertion loss of ~5 dB (averaged over all the optical paths). All the 132 tunable elements of the processor provide more than $2\pi$ phase shift with high extinction ratio. High fidelities are measured over a set of 1036 unitary transformations. Quantum interference of high visibility is replicable across the entire processor, i.e., the indistinguishability of photons is preserved.

The photonic processor presented here is the largest low-loss plug-and-play universal square photonic processor to date, enabling fully-reconfigurable unitary transformations across 12 inputs and through 12 layers of depth.

**References**

1. Carolan, J. *et al.* Universal linear optics. *Science* **349**, 711 (2015).
2. Qiang, X. *et al.* Large-scale silicon quantum photonics implementing arbitrary two-qubit processing. *Nat. Photonics* **12**, 534–539 (2018).
3. Santagati, R. *et al.* Silicon photonic processor of two-qubit entangling quantum logic. *J. Opt.* **19**, 114006 (2017).
4. Taballione, C. *et al.* 8×8 reconfigurable quantum photonic processor based on silicon nitride waveguides. *Opt. Express* **27**, 26842–26857 (2019).
5. Ribeiro, A., Ruocco, A., Vanacker, L. & Bogaerts, W. Demonstration of a 4 × 4-port universal linear circuit. *Optica* **3**, 1348–1357 (2016).
6. Koteva, K. I. *et al.* Silicon quantum photonic device for multidimensional controlled unitaries. in *Frontiers in Optics / Laser Science* FTu8D.1 (Optical Society of America, 2020).
7. Harris, N. C. *et al.* Quantum transport simulations in a programmable nanophotonic processor. *Nat. Photonics* **11**, 447–452 (2017).
8. Sparrow, C. *et al.* Simulating the vibrational quantum dynamics of molecules using photonics. *Nature* **557**, 660–667 (2018).
9. Carolan, J. *et al.* Variational quantum unsampling on a quantum photonic processor. *Nat. Phys.* **16**, 322–327 (2020).
10. Spring, J. B. *et al.* Boson Sampling on a Photonic Chip. *Science* **339**, 798 (2013).
11. Tillmann, M. *et al.* Experimental boson sampling. *Nat. Photonics* **7**, 540–544 (2013).
12. Shadbolt, P. J. *et al.* Generating, manipulating and measuring entanglement and mixture with a reconfigurable photonic circuit. *Nat. Photonics* **6**, 45–49 (2012).
13. Paesani, S. *et al.* Generation and sampling of quantum states of light in a silicon chip. *Nat. Phys.* **15**, 925–929 (2019).
14. Lee, Y., Bersin, E., Dahlberg, A., Wehner, S. & Englund, D. A Quantum Router Architecture for High-Fidelity Entanglement Flows in Quantum Networks. *arXiv:2005.01852* (2020).
15. Chen, K. C., Bersin, E. & Englund, D. A Polarization Encoded Photon-to-Spin Interface. *arXiv:2004.02381* (2020).
16. Wan, N. H. *et al.* Large-scale integration of artificial atoms in hybrid photonic circuits. *Nature* **583**, 226–231 (2020).
17. Choi, H., Pant, M., Guha, S. & Englund, D. Percolation based architecture for cluster state creation using photon-mediated entanglement between atomic memories. *arXiv:1704.07292* (2019).
18. Steinbrecher, G. R., Olson, J. P., Englund, D. & Carolan, J. Quantum optical neural networks. *Npj Quantum Inf.* **5**, 60 (2019).
19. Shen, Y. *et al.* Deep learning with coherent nanophotonic circuits. *Nat. Photonics* **11**, 441–446 (2017).
20. Gentile, A. A. *et al.* Learning models of quantum systems from experiments. *arXiv:2002.06169* (2020).
21. Zhuang, L., Roeloffzen, C. G. H., Hoekman, M., Boller, K.-J. & Lowery, A. J. Programmable photonic signal processor chip for radiofrequency applications. *Optica* **2**, 854–859 (2015).
22. Pérez, D. *et al.* Multipurpose silicon photonics signal processor core. *Nat. Commun.* **8**, 636 (2017).
23. Reck, M., Zeilinger, A., Bernstein, H. J. & Bertani, P. Experimental realization of any discrete unitary operator. *Phys. Rev. Lett.* **73**, 58–61 (1994).
24. Clements, W. R., Humphreys, P. C., Metcalf, B. J., Kolthammer, W. S. & Walmsley, I. A. Optimal design for universal multiport interferometers. *Optica* **3**, 1460–1465 (2016).
25. O'Brien, J. L., Furusawa, A. & Vučković,

J. Photonic quantum technologies. *Nat. Photonics* **3**, 687–695 (2009).

26. Bromley, T. R. *et al.* Applications of Near-Term Photonic Quantum Computers: Software and Algorithms. *arXiv:1912.07634* (2019).

27. Kok, P. *et al.* Linear optical quantum computing with photonic qubits. *Rev. Mod. Phys.* **79**, 135–174 (2007).

28. Zhong, H.-S. *et al.* Quantum computational advantage using photons. *Science* eabe8770 (2020) doi:10.1126/science.abe8770.

29. C. G. H. Roeloffzen *et al.* Low-Loss Si3N4 TriPleX Optical Waveguides: Technology and Applications Overview. *IEEE J. Sel. Top. Quantum Electron.* **24**, 1–21 (2018).

30. Skaar, J., García Escartín, J. C. & Landro, H. Quantum mechanical description of linear optics. *Am. J. Phys.* **72**, 1385–1391 (2004).

31. Hong, C. K., Ou, Z. Y. & Mandel, L. Measurement of subpicosecond time intervals between two photons by interference. *Phys. Rev. Lett.* **59**, 2044–2046 (1987).

32. Donzella, V., Talebi Fard, S. & Chrostowski, L. Study of waveguide crosstalk in silicon photonics integrated circuits. *Proc. SPIE* 8915, *Photonics North* 2013, 89150Z (2013).

33. Ceccarelli, F. *et al.* Low Power Reconfigurability and Reduced Crosstalk in Integrated Photonic Circuits Fabricated by Femtosecond Laser Micromachining. *Laser Photonics Rev.* **14**, 2000024 (2020).

34. M. Milanizadeh, D. Aguiar, A. Melloni & F. Morichetti. Canceling Thermal Cross-Talk Effects in Photonic Integrated Circuits. *J. Light. Technol.* **37**, 1325–1332 (2019).

35. Miller, D. A. B. Perfect optics with imperfect components. *Optica* **2**, 747–750 (2015).

36. Burgwal, R. *et al.* Using an imperfect photonic network to implement random unitaries. *Opt. Express* **25**, 28236–28245 (2017).

37. Zhao, Y. *et al.* Microresonator Based Discrete- and Continuous-Variable Quantum Sources on Silicon-Nitride. *OSA Quantum 20 Conf.* **QM4B.3**, (2020).

38. Schuck, C. *et al.* Quantum interference in heterogeneous superconducting-photonic circuits on a silicon chip. *Nat. Commun.* **7**, 10352 (2016).

39. Schuck, C., Pernice, W. H. P. & Tang, H. X. NbTiN superconducting nanowire detectors for visible and telecom wavelengths single photon counting on Si3N4 photonic circuits. *Appl. Phys. Lett.* **102**, 051101 (2013).

40. K. Wörhoff *et al.* Liquid crystal phase modulator integration on the TriPleX photonic platform. *Proc. SPIE* 11180, *International Conference on Space Optics* – ICSO 2018, 1118077 (2019).

41. Cano-Garcia, M. *et al.* Integrated Mach–Zehnder Interferometer Based on Liquid Crystal Evanescent Field Tuning. *Crystals* **9**, 225 (2019).

42. Ríos, C. *et al.* Integrated all-photonic non-volatile multi-level memory. *Nat. Photonics* **9**, 725–732 (2015).

43. Hosseini, N. *et al.* Stress-optic modulator in TriPleX platform using a piezoelectric lead zirconate titanate (PZT) thin film. *Opt. Express* **23**, 14018–14026 (2015).

44. Epping, J. P. *et al.* Ultra-low-power stress-based phase actuator for microwave photonics. *2017 Eur. Conf. Lasers Electro-Opt. Eur. Quantum Electron. Conf.* **CK_7_6**, (2017).

45. Stanfield, P. R., Leenheer, A. J., Michael, C. P., Sims, R. & Eichenfield, M. CMOS-compatible, piezo-optomechanically tunable photonics for visible wavelengths and cryogenic temperatures. *Opt. Express* **27**, 28588–28605 (2019).

46. Uppu, R. *et al.* On-chip deterministic operation of quantum dots in dual-mode waveguides for a plug-and-play single-

photon source. *Nat. Commun.* **11**, 3782 (2020).

47. Pérez, D., Gutierrez, A. M., Sanchez, E., DasMahapatra, P. & Capmany, J. Integrated photonic tunable basic units using dual-drive directional couplers. *Opt. Express* **27**, 38071–38086 (2019).

48. N. Quack *et al*., MEMS-Enabled Silicon Photonic Integrated Devices and Circuits. *IEEE Journal of Quantum Electronics* **56**, 1, 1-10 (2020).

---

[i] Throughout this paper, 'modes' refers to the zeroth order mode of each waveguide. We note that each waveguide supports multiple frequency modes.